# Abnormal anti-crossing effect in photon-magnon coupling

Biswanath Bhoi, Bosung Kim, Seung-Hun Jang, Junhoe Kim, Jaehak Yang, Young-Jun Cho, and Sang-Koog Kim[a]),

*National Creative Research Initiative Center for Spin Dynamics and Spin-Wave Devices, Nanospinics Laboratory, Research Institute of Advanced Materials, Department of Materials Science and Engineering, Seoul National University, Seoul 151-744, Republic of Korea*

We report the experimental demonstration of an abnormal, opposite anti-crossing effect in a photon-magnon-coupled system that consists of an Yttrium Iron Garnet film and an inverted pattern of split-ring resonator structure (noted as ISRR) in a planar geometry. It is found that the normal shape of anti-crossing dispersion typically observed in photon-magnon coupling is changed to its opposite anti-crossing shape just by changing the position/orientation of the ISRR's split gap with respect to the microstrip line axis along which ac microwave currents are applied. Characteristic features of the opposite anti-crossing dispersion and its linewidth evolution are analyzed with the help of analytical derivations based on electromagnetic interactions. The observed opposite anti-crossing dispersion is ascribed to the compensation of both intrinsic damping and coupling-induced damping in the magnon modes. This compensation is achievable by controlling the relative strength and phase of oscillating magnetic fields generated from the ISRR's split gap and the microstrip feeding line. The position/orientation of an ISRR's split gap provides a robust means of controlling the dispersion shape of anti-crossing and its damping in a photon-magnon coupling, thereby offering more opportunity for advanced designs of microwave devices.

[a]) Correspondence and requests for materials should be addressed to S.-K. K

sangkoog@snu.ac.kr

# I. INTRODUCTION

Understanding and exploiting the interactions of excited modes in hybrid quantum systems are the key to building large-scale artificial many-body quantum systems such as quantum computers, quantum communication networks, and quantum simulators [1-3]. Therefore, collectively excited modes (i.e., magnons) in ferromagnets or ferrimagnets, being coupled to elementary excitations of electromagnetic waves (photons), have increasingly been studied in a variety of hybrid structures of two or more different systems [4-12]. In particular, the rapid development of both spintronics and the design/fabrication technologies of microwave resonators has stimulated further studies of photon-magnon coupling using a low-damping magnetic material, e.g., yttrium iron garnet (YIG: $Y_3Fe_5O_{12}$), and high-quality microwave resonators [13-21]. In earlier studies, the interaction (coupling) between the photon and magnon modes usually has been demonstrated by showing the modes' splitting at and near their common resonant frequency within the so-called anti-crossing or the level repulsion of two coupled modes [13-21]. The energy split-gap in such anti-crossing increases with the modes' coupling strength, as described well by a classical model for coupled oscillators [13,14,16-19,21] and also by a two-level quantum model [7-16].

The level attraction has been theoretically predicted in coupled modes [22-28]. If the off-diagonal term of the Hamiltonian of a coupled system contains an imaginary part, the eigenfrequency of the coupled modes becomes complex and the corresponding real component pulls towards each other to meet. This level attraction can be achieved using coupled oscillators of positive and negative energy [22-24], as recently demonstrated successfully in quantum optomechanical systems [27, 29] at a temperature of less than 50 mK using a complicated experimental setup wherein a nano-mechanical oscillator is dispersively coupled to a driven optical cavity [27, 29]. Similar anti-crossing behavior has been observed in plasmonic nanostructures due to either near-field or far-field coupling [30-32]. An alternative example is coupling between two modes of a non-Hermitian Hamiltonian, which interaction is also referred to as dissipative coupling (or external-mode coupling) [27-28, 30-31]. Although these effects are potentially applicable in the fields of topological energy transfer, quantum sensing, and nonreciprocal photon transmission [28, 29, 33], they have yet to be explored in photon-magnon-coupled systems. Despite Grigoryan *et al.*'s [34] report of a theoretical framework for observation of such level attraction in a spin-photon system, this phenomenon has yet to be found experimentally, other than in the recent study of Harder *et al.* [28].

Herein we report an experimental demonstration of abnormal, opposite anti-crossing phenomenon in a photon-magnon-coupled system that consists of an inverted pattern of split ring resonator (hereafter noted as ISRR) and a YIG film. The planar, hybrid ISRR-YIG system is more suitable for practical applications than are three-dimensional (3-D) complex systems [35]. This work is not only of fundamental interest with respect to the nonlinear phenomenon of dissipative quantum systems but also offers a platform for exploring the underlying physics of coupling in a variety of hybrid systems including magnon-phonon, plasmon-magnon, and exciton-photon coupling [9,20].

## II. EXPERIMENTAL DETAILS

The present experimental setup for photon-magnon coupling measurements of the ISRR-YIG hybrid is shown in Fig. 1, wherein the ISRR on the ground plane (dark yellow) excites a microwave photon mode to be coupled to the magnon modes in the YIG film (green color). The dimensions of the ISRR and the YIG film are exactly the same as those reported earlier [18]. We designed and fabricated two different samples: the split-gap position of the ISRR was placed on the x-axis (case-I) and the y-axis (case-II), as shown in the inset of Fig. 1.

The details on the fabrication of the ISRRs and the measurement of photon-magnon coupling are available in Ref. [18].

## III. RESULTS AND DISCUSSION

From the two different orientations of the ISRR's split gap in the ISRR-YIG hybrid, we experimentally measured the transmission coefficient $S_{21}$ spectra as a function of the microwave frequency ($f = \omega/2\pi$) of oscillating currents flowing along the microstrip line (on the *y*-axis) for different strengths *H* of static magnetic fields applied along the *x*-axis. From the observed spectra, we replotted the $S_{21}$ power on the $\omega/\omega_{ISRR} - H/H_{cent}$ plane, as shown in Fig. 2, where $\omega_{ISRR}$ and $H_{cent}$ represent the resonant angular frequency of the ISRR and the field center ($\omega_r = \omega_{ISRR}$) of the anti-crossing dispersion, respectively. Owing to a strong coupling between the YIG's magnon and the ISRR's photon modes, for case-I, the normal shape of anti-crossing dispersion was found, which represents two higher and lower coupled branches around the resonant frequencies of the FMR mode and the photon mode (see Fig. 2(a)). On the other hand, for case-II, an abnormal anti-crossing dispersion was observed, as shown in Fig. 2(b). Quite recently, such abnormal anti-crossing dispersion was also experimentally observed, but in complex cavity optomechanical systems at a very low temperature below 50 mK [27].

To quantitatively analyze the similarity and difference of such experimentally observed normal and abnormal dispersion shapes that vary only with the position/orientation of the ISRR's split gap, we made analytical derivations based on electromagnetic interactions (Faraday's induction and Ampere's circuit laws) between the YIG film and the ISRR. The hybrid system used in this study consists of three different physical systems: 1) the microstrip line to excite the magnon and photon modes as well as to probe those coupled modes; 2) the YIG, and 3) the ISRR wherein the magnon and photon modes are to be excited respectively. In our analytical derivations, we considered all the three interactions, namely between ① and ②, ① and ③, and ② and ③. The ISRR was lying on the *x*-*y* plane, and ac currents were applied along the microstrip line placed on the *y*-axis. For only the ISRR, the ac currents can excite via both Ampere's circuit and Faraday's induction laws. Consequently, the ac current *j* in the microstrip line yields an electromotive force (EMF) voltage *V* in the ISRR, as expressed by $V = Z_{ISRR} j$, $Z_{ISRR}$ being the ISRR's impedance, which is given, according to an equivalent *LCR* circuit model, as

$$Z_{ISRR} = -\frac{iL}{\omega}\left(\omega^2 - \omega_{ISRR}^2 + 2i\beta\omega\omega_{ISRR}\right), \quad (1)$$

where $\omega_{ISRR} = 1/\sqrt{LC}$ is the angular resonance frequency of ISRR with inductance *L* and capacitance *C*, and $\beta = R/2L\omega_{ISRR}$ is the damping parameter with resistance *R*. According to the relation of $\beta = \Delta\omega_{HWHM}/\omega_{ISRR}$ with the half width at half maximum $\Delta\omega_{HWHM}$ of $|S_{21}|$ spectra [12,15,19], $\beta$ was estimated to be $\sim 2.0 \times 10^{-2}$ using experimentally observed data from only the ISRR without a YIG film (see Supplementary S2).

For the YIG only, the oscillating magnetic field created (via Ampere's circuit law) by the ac currents flowing along the microstrip line can directly stimulate magnetization excitations in the YIG film, as described by the Landau-Lifshitz-Gilbert (LLG) equation [28,34]

$$\frac{d\mathbf{m}}{dt} = -\gamma\mathbf{m}\times\mathbf{H}_{eff} + \alpha\mathbf{m}\times\frac{d\mathbf{m}}{dt}, \quad (2)$$

where $\mathbf{m} = \mathbf{M}/M_s$ is the magnetization vector, with gyromagnetic ratio $\gamma/2\pi$ = 28 GHz/T,

intrinsic Gilbert damping parameter $\alpha = 3.2 \times 10^{-4}$ and saturation magnetization $\mu_0 M_s =$ 0.172 T, as derived from the FMR measurement of only the YIG film. $\mathbf{H}_{eff}$ is the effective magnetic field, given as $\mathbf{H}_{eff} = \mathbf{H} + \mathbf{h}_{line}$, where $\mathbf{H} = H\hat{\mathbf{x}}$ is the static magnetic field externally applied in the *x*-direction and $\mathbf{h}_{line} = \mathbf{h}e^{-i\omega t}$ is the ac magnetic field generated from the microstrip line with amplitude $|\mathbf{h}|$ and angular frequency $\omega$. Using a linearized form of the magnetization direction, the magnetization variation is given as $\mathbf{m} \cong M_s\hat{\mathbf{x}} + \mathbf{m}_\perp e^{-i\omega t}$, where $\mathbf{m}_\perp e^{-i\omega t}$ is the oscillating component of the magnetization on the *y-z* plane. Assuming $|\mathbf{m}_\perp| \ll M_s$, the LLG equation can be simplified in the rotational frame to [10,16,34]

$$(\omega - \omega_r + i\alpha\omega)m^+ + \omega_m h^+ = 0 , \qquad (3)$$

where $m^+ = m_y + im_z$, $h^+ = h_y + ih_z$, and the FMR resonance frequency $\omega_r = \gamma\sqrt{H(H + \mu_0 M_s)}$, $\omega_m = \gamma\mu_0 M_s$ [12,12,16,34].

Next, let us consider the interaction between the YIG and the ISRR modes. The two modes mutually interact when excited. Therefore, in the derivations, we take into account the law of action-reaction. Once the magnetizations are excited in the YIG, they can yield an additional voltage to the ISRR according to Faraday's induction law, as given by $V_y = K_F L(dm_z/dt)$ and $V_z = -K_F L(dm_y/dt)$. The total induced voltage in the ISRR is thus $V_{ISRR \leftarrow YIG} = V_y + iV_z = -K_F L\omega m^+$, where $K_F$ is the coupling parameter to account for the phase relation between the ISRR's photon and the YIG's magnon modes according to Faraday's induction law. This induced voltage generates an additional microwave current in the ISRR, as expressed by $V_{ISRR \leftarrow YIG} = Z_{ISRR} J^+$, where $J^+ \equiv J_y + iJ_z = Je^{-i(\omega t + \phi)}$ is the net current in the ISRR circuit, and $\phi$ is the phase difference between the currents in the microstrip line and the ISRR, respectively. Using Eq. (1), this relation is finally written as

$$iK_F\omega^2 m^+ + \left(\omega^2 - \omega_{ISRR}^2 + 2i\beta\omega\omega_{ISRR}\right)J^+ = 0 , \qquad (4)$$

Additionally, according to the action and reaction law, the ISRR's induced currents also create a strong microwave magnetic field around the ISRRs' split gap. This field, in turn, contributes to the excitation of magnetizations in the YIG [16, 28, 34]. Thus, the magnetizations in the YIG are influenced by the effective field, which is the sum of two time-dependent magnetic fields $\mathbf{h}_{line}\left(=\mathbf{h}e^{-i\omega t}\right)$ from the feeding line and $\mathbf{h}_{ISRR}\left(=\delta\mathbf{h}e^{-i(\omega t+\phi)}\right)$ from the ISSR split gap, where $\phi$ is the phase difference between the two fields $\mathbf{h}_{ISRR}$ and $\mathbf{h}_{line}$. Here, we assume that the magnitude of $\mathbf{h}_{ISRR} = \delta e^{-i\phi}\mathbf{h}_{line}$, with $\delta = |\mathbf{h}_{ISRR}|/|\mathbf{h}_{line}|$. Here too, both the values of $\phi$ and $\delta$ can be controlled by the ISRR's split-gap orientation with respect to the microstrip line axis. Taking into account the total magnetic field that contributes to the magnetization excitation in the YIG film, we have $\mathbf{H}_{eff} = \mathbf{H} + \mathbf{h}_{line} + \mathbf{h}_{ISRR} = H\hat{\mathbf{x}} + \left(1+\delta e^{-i\phi}\right)\mathbf{h}e^{-i\omega t}$. Finally, the LLG equation (Eq. 3) in the rotating frame is thus rewritten as

$$\left(\omega - \omega_r + i\alpha\omega\right)m^+ - i\omega_m K_A\left(1+\delta e^{i\phi}\right)J^+ = 0 , \qquad (5)$$

where $J^+ = i\left(h^+\right)_{ISRR}/K_A$ is the net microwave current in the ISRR circuit, resulting in the magnetic field of $h_{ISRR}$ via Ampere's law, as $\left(h_y\right)_{ISRR} = K_A J_z$ and $\left(h_z\right)_{ISRR} = -K_A J_y$, and $K_A$ is the coupling parameter that determines the phase relation between the ISRR photon and YIG magnon modes due to Ampere's law. To obtain the simultaneous solutions of Eqs. (4) and (5), the matrix form is rewritten as

$$\begin{pmatrix} \omega^2 - \omega_{ISRR}^2 + 2i\beta\omega\omega_{ISRR} & iK_F\omega^2 \\ -i\omega_m\left(1+\delta e^{i\phi}\right)K_A & \omega - \omega_r + i\alpha\omega \end{pmatrix} \begin{pmatrix} J^+ \\ m^+ \end{pmatrix} = \begin{pmatrix} 0 \\ 0 \end{pmatrix}, \quad (6a)$$

$$\Omega \begin{pmatrix} m^+ \\ J^+ \end{pmatrix} = \begin{pmatrix} 0 \\ 0 \end{pmatrix}, \quad (6b)$$

The determinant of $\Omega$ is expressed as $\left(\omega - \omega_r + i\alpha\omega_r\right)\left(\omega^2 - \omega_{ISRR}^2 + 2i\beta\omega\omega_{ISRR}\right) - K_A K_F \omega_m \omega^2 \left(1 + \delta e^{i\phi}\right) = 0$, $K^2 \cong K_A K_F$; as such, it finally describes photon-magnon coupling in the ISRR-YIG hybrid system. The coupling of the magnetizations in the YIG to currents in the ISRR is represented by the matrix's first line, which describes the LRC circuit of the ISRR affected by the magnetization motions of the YIG film, while the effect of the net currents of the ISRR on the magnetization dynamics in the YIG film is described by the second line of the matrix. Here, the phase difference $\phi$ between the currents flowing in the microstrip line and the ISRR gives rise to the effective microwave magnetic field that excites magnons in the YIG film. The two parameters of $\phi$ and $\delta$ can be readily controlled by changing the position/orientation of the ISRR's split gap with respect to the microstrip line axis, which was confirmed by the electromagnetic simulation of the ISRRs for two different split gap orientations without the YIG film (see Supplementary S1). For the case of the orientation of the ISRRs' split gap being perpendicular to the microstrip line (case-I), the microwave fields of both $\mathbf{h}_{ISRR}$ and $\mathbf{h}_{line}$ are in the equal phase. On the other hand, for the case of the orientation of the ISRRs' split gap being parallel to the microstrip line (case-II), the microwave fields of both $\mathbf{h}_{ISRR}$ and $\mathbf{h}_{line}$ are out of ($\pi$) phase [34, 36-38].

In order to understand such contrasting anti-crossing dispersions experimentally observed for the case I and case II geometries, we formulated a simple analytical expression for both dispersion curves in an anti-crossing region in our experimental case of $\alpha = 3.2 \times 10^{-4}, \beta = 2.0 \times 10^{-2} \ll 1$, where the approximate solution of the real part of $\Omega = 0$ becomes (see Supplementary S4)

$$\omega_\pm \approx \frac{1}{2}\left[(\omega_r + \omega_{ISRR}) \pm \sqrt{(\omega_r - \omega_{ISRR})^2 + (4\pi\Delta)^2}\right], \tag{7a}$$

$$\Delta = \frac{1}{4\pi}\sqrt{2K^2\omega_m\omega_{ISRR}(1 + \delta\cos\phi)}, \tag{7b}$$

where $\Delta$ is the net coupling strength at $H = H_{cent}$ ($\omega_r = \omega_{ISRR}$) in Hz units. From the fitting of Eq. (7) to the lower and higher frequency branches (black solid lines) shown in Fig. 2, we obtained the fitting values of $\Delta = 90$ and $90i$ MHz for the normal (Fig. 2(a)) and the opposite (Fig. 2(b) anti-crossing dispersions, respectively. The experimental results are well fitted with Eq. (7). We note that for the opposite anti-crossing, we obtained only the imaginary value of $\Delta$, whereas for the normal anti-crossing, the real value of $\Delta$. Therefore, the real and imaginary values for $\Delta$ characterize the normal and the opposite anti-crossing dispersions, respectively. According to Eq. (7b), the real and imaginary numbers of $\Delta$ are the cases of $(1 + \delta\cos\phi) > 0$ and $(1 + \delta\cos\phi) < 0$, respectively. Thus, in the next paragraph, we will discuss how $\delta$ and $\phi$ affect the type of anti-crossing dispersion.

In order to examine how $\delta$ and $\phi$ determine anti-crossing behaviors, from Eq. (6) we further numerically calculated the complex eigenvalues of two coupled modes, i.e., $E_\pm = \omega_\pm - i\Delta\omega_\pm$, where $\omega_\pm$ and $\Delta\omega_\pm$ represent the dispersion shape and the linewidth evolution of the coupled modes, respectively [7,10,12,16]. In this numerical calculation, we used constant values of $\alpha = 3.2 \times 10^{-4}$ and $\beta = 0.02$, $K = 0.03$, and $\omega_{ISRR}/2\pi = 3.7$ GHz (for case-I) and 4.1 GHz (for case-II), all of which values were experimentally estimated, as mentioned earlier and in Suppl. S5. For two specific cases of $(\delta, \phi) = (1/2, 0)$ and $(2, \pi)$, the resultant numerical calculations are given in Figs. 3(a) and 3(b), respectively. For ($\delta$, $\phi$) = (1/2, 0), the $\omega_+$ and $\omega_-$ branches repel each other (top of Fig. 3(a)) while their linewidths $\Delta\omega_+$ and $\Delta\omega_-$ cross each other at $H_{cent}$ (bottom of Fig. 3(a)). This behavior is ubiquitous from solid-state theory to quantum chemistry when certain energies are transferred mutually (reciprocally) between one and the other in a variety of coupled systems as reported in Refs. [23, 39-40]. On the other hand, for the case of ($\delta$, $\phi$) = (2, $\pi$), $\omega_+$ and $\omega_-$ provide the energies of their states that attract each other and nearly meet at two points noted as P1 and P2, where the curves have kinks (marked by the vertical dotted lines in Fig. 3(b)), while $\Delta\omega_+$ and $\Delta\omega_-$ are found to be repulsive (they do not cross each other), as shown in bottom of Fig. 3(b).

Interestingly, the real parts $\omega_+$ and $\omega_-$ of the calculation results for two cases of ($\delta$, $\phi$) = (1/2, 0) and (2, $\pi$) are similar to the experimentally observed dispersion spectra (Fig. 2) for the case-I and case-II geometries of ISRR's split gap, respectively. When $\phi = 0$, $\delta < 1$ ($|\mathbf{h}_{ISRR}| < |\mathbf{h}_{line}|$ for our case-I), the energy exchange takes place between the ISRR photon and the YIG magnon modes. This condition usually results in the modification of damping in both modes: lower damping in the magnon mode and higher damping in the photon mode, as shown in the bottom of Fig. 3(b). In the center of the anti-crossing region ($H_{cent}$), $\Delta\omega_+$ and $\Delta\omega_-$ cross each other, representing an equal $\Delta\omega$ value. Here, the transmission coefficient including the dispersion and the linewidth depends on the interplay between the damping of the two interacting modes and the coupling strength [16, 40].

On the other, when $\delta > 1$ and $\phi = \pi$ ($|\mathbf{h}_{ISRR}| > |\mathbf{h}_{line}|$ for our case-II), the second driving force $\mathbf{h}_{ISRR}$ exerts an anti-damping torque that compensates the intrinsic damping and

coupling-induced damping in the magnon modes. As a result, the magnon mode absorbs energy and acts as a pumping source for the ISRR photon mode [34]. It can be stated that if the phase of $\mathbf{h}_{ISRR}$ becomes deviated to that of $\mathbf{h}_{line}$, $\mathbf{h}_{ISRR}$ contributes as negative damping (decreased damping) to the magnon mode and as positive damping to the photon mode. Therefore, linewidth bifurcation starts near the point P1, reaches the maximum in the middle between the two P1 and P2 (bottom of Fig. 3(b)), and ends near the P2, beyond which the linewidths of both states are almost equal to their initial state. This is a signature of purely dissipative (non-Hermitian) interaction of two coupled modes where one of the modes has a larger loss rate than does the other [30-32]. The presence of this kind of dissipative coupling between the two modes results in an increase of the lifetime of the magnon mode and a decrease of the lifetime of the photon mode [30-31, 28, 34], as shown by the relatively large difference between $\Delta\omega_+$ and $\Delta\omega_-$ around the center field. The strong coupling of low-damping magnons to the photon together with a large-phase-shifted driving force makes the linewidth of the magnons negative for the opposite anti-crossing dispersion [34]. Similar to our present results, such linewidth bifurcation was observed in coupled systems during level attraction [23, 28, 39].

Next, in order to further examine the anti-crossing effect versus $\delta$ and $\phi$, we numerically calculated |$S_{21}$| using input-output formalism [10,12,16,19, 34], as given by

$$S_{21} = \Gamma \frac{J^+}{j} = \Gamma \frac{\omega^2(\omega - \omega_r + i\alpha\omega)}{\det(\Omega)} , \qquad (8)$$

where $j$ is the ac current of the microstrip line and $\Gamma \approx 2\beta$ for the ISRR-YIG hybrid/cable impedance mismatch [7, 34]. Then, using Eq. (8), numerical calculations of |$S_{21}$| power on the $\omega/\omega_{ISRR} - H/H_{cent}$ plane were performed by varying both $\delta$ and $\phi$ values. Figure 4(a) shows

the calculated |S$_{21}$| profiles versus $\omega/\omega_{ISRR}$ at the center position ($H_{cent}$). Figure 4(b) shows many contrasting shapes of anti-crossing dispersion depending on the indicated values of the $\delta$ and $\phi$ rather than the appearance of only the normal and the opposite anti-crossing dispersions experimentally found. The distinct features can be categorized into three types: (i) normal anti-crossing, (ii) opposite anti-crossing, and (iii) non-anti-crossing. The shape of dispersion is totally determined by the net coupling strength $\Delta$, which values vary with $\delta$ and $\phi$, as shown in Fig. 4(c). For example, for $\delta = 0$, we have $\Delta = \frac{1}{4\pi}\sqrt{2K^2\omega_m\omega_{ISRR}}$; thus the variation of $\phi$ does not affect $\Delta$ nor the shape of anti-crossing dispersion; all of the |S$_{21}$| power contours show the normal type of anti-crossing, as indicated in the first rows of Figs. 4(a) and 4(b). This condition corresponds to the cases of $|\mathbf{h}_{ISRR}| \ll |\mathbf{h}_{line}|$ in our experiments. On the other hand, for the case of $\delta = 1$, Eq. (7) is a function of $\phi$; thus the shape of anti-crossing is remarkably variable with $\phi$. However, under a specific condition of $\phi = \pi$, $\Delta$ becomes zero, thus the anti-crossing dispersion completely disappears. For the case of $\delta = 1$ and $\phi = 0$, in our system $\mathbf{h}_{line}$ and $\mathbf{h}_{ISRR}$ are comparable in size and equal in phase, and thus, both fields excite the YIG's magnon modes. As $\phi$ increases from 0 to $\pi$, $\mathbf{h}_{ISRR}$ becomes more out-of-phase with $\mathbf{h}_{line}$, thereby yielding weaker net coupling strength. For the condition of $\delta = 1$, $\phi = \pi$, $\mathbf{h}_{ISRR}$ and $\mathbf{h}_{line}$ are exactly out-of-phase, thus yielding a completely zero microwave field, which cannot excite YIG's magnons, as shown by the appearance of only ISRR's photon mode, without the FMR mode in YIG (⑩ of Figs. 4(a) and 4(b)).

More interestingly, for the case of $\delta = 2$, the anti-crossing shape changes from the normal to the opposite one through non-anti-crossing (see the third rows in Figs. 4(a) and 4(b)). The |$\Delta$| value decreases with $\phi$ and becomes 0 for $\phi = 2\pi/3$ and increases again with

$\phi$ from $\phi = 2\pi/3$ (see the bottom of the right column in Fig. 4(b)). For the case of $\delta = 2$, $\phi = 0$, the YIG magnon mode is mainly excited by the $\mathbf{h}_{ISRR}$, and the $\mathbf{h}_{ISRR}$ and $\mathbf{h}_{line}$ are in-phase, resulting in the normal shape of anti-crossing. At $\phi = 2\pi/3$, $|\Delta|$ becomes zero and the anti-crossing disappears at the common resonant peak. For $\phi > 2\pi/3$, $|\Delta|$ increases with $\phi$, resulting in the opposite anti-crossing. With increasing $|\Delta|$, the opposite anti-crossing becomes clearer in its shape. Since $\mathbf{h}_{ISRR}$ contributes more in magnon excitations than $\mathbf{h}_{line}$ does and the fields are out-of-phase, the result is the opposite anti-crossing. All of these features clearly indicate that the relative strength and phase of the oscillating magnetic fields generated from both the ISRR's split gap and the microstrip feeding line determine the net coupling strength, consequently resulting in the shape of anti-crossing dispersion in the ISRR-YIG hybrid system.

Furthermore, a phase diagram of anti-crossing dispersion on the plane of $\phi$ and $\delta$ is calculated using $K = 0.03$ and $\omega_{ISRR}/2\pi = 3.7$ GHz according to $\Delta = \frac{1}{4\pi}\sqrt{2K^2\omega_m\omega_{ISRR}(1+\delta\cos\phi)}$ as shown in Fig. 4(c). The opposite anti-crossing dispersions (blue region) are separated from the others (the red and yellow regions) by the condition of $\Delta = 0$ (i.e., $\delta\cos\phi = -1$), as indicated by the black solid line in Fig. 4(c). As noted by the colors in the different regions, as $\delta$ increases in the range of $\phi < \pi/2$, the anti-crossing becomes normal with stronger net coupling strength, whereas as $\delta$ increases in the range of $\phi > 3\pi/4$ above the marked boundary curve, the anti-crossing becomes the opposite with stronger net coupling strength.

In earlier studies on non-Hermitian and optomechanical systems, it was shown that the presence of damping might affect the anti-crossing dispersion [27, 40]. The eigenvalues of non-Hermitian Hamiltonian can show the crossing (opposite anti-crossing) or anti-crossing type of dispersion for any coupling strength, depending on the difference in damping between two different systems [27, 40]. To explore the role of each damping in a strongly coupled photon-magnon system and how each damping together with $\phi$ and $\delta$ affects the net coupling strength, we derived a generalized analytical form for net coupling strength using Eq. (6) (for details, see Supplementary S7) as given by

$$\Delta' = \frac{\sqrt{2}K}{4\pi}\sqrt{\omega_{ISRR}\omega_m\left(1+\delta\cos\phi\right) - \omega_{ISRR}^2\left(\beta-\alpha\right)^2/2K^2} \tag{9}$$

The real and imaginary value of $\Delta'$ is determined by the sign of $\omega_m\omega_{ISRR}\left(1+\delta\cos\phi\right) - \omega_{ISRR}^2\left(\beta-\alpha\right)^2/2K^2$, thus approximately by the relative magnitude of the intrinsic parameter $\Delta_{mat} = (\beta-\alpha)^2/2K^2$ and the geometry parameter $\Delta_{geom} = \left(1+\delta\cos\phi\right)$. For the case of $\Delta_{mat} = (\beta-\alpha)^2/2K^2 = 0$, Eq. (9) becomes Eq. (7b). We note that the color-bar scale in Fig. 4 (c) represents the net coupling strength only for the case of $\Delta_{mat} = (\beta-\alpha)^2/2K^2 = 0$. For our case with $\beta = 2.0\times10^{-2}$, $\alpha = 3.2\times10^{-4}$ and $K = 0.03$, the condition of $\Delta' = 0$ is also drawn as a red solid line on the phase diagram, as shown in Fig. 4(c). The boundary curve is close to that of $\Delta = 0$. To examine how $\Delta'$ varies with $\Delta_{mat}$, we plotted the phase diagram for $\Delta_{mat}$ = 0.1, 0.3, 0.5, and 0.9 in Supplementary S8. Since $\Delta' = 0$ represents the boundary that distinguishes the opposite anti-crossing dispersion from the others, the phase diagram on the plane of $\phi$ and $\delta$ is also modified according to the value of $\Delta_{mat}$. Here, what is most important is that with increasing $\Delta_{mat}$, the opposite anti-crossing region expands towards lower values of both $\phi$ and $\delta$. As we discussed earlier when the phase of $\mathbf{h}_{ISRR}$ becomes deviated to that of $\mathbf{h}_{line}$, the damping of the photon mode increases while the damping of the magnon mode decreases due to dissipative interaction, thus leading to an opposite anti-crossing dispersion. For larger $\Delta_{mat}$, the opposite anti-crossing dispersion occurs for the smaller phase difference between $\mathbf{h}_{ISRR}$ and $\mathbf{h}_{line}$ and the lower value of $\delta$ ( $=|\mathbf{h}_{ISRR}|/|\mathbf{h}_{line}|$ ). In the case of $\Delta_{geom}=(1+\delta\cos\phi) \gg \Delta_{mat}=(\beta-\alpha)^2/2K^2$, the anti-crossing dispersion is not much varied with $\alpha$ and $\beta$, but rather is determined dominantly by $\phi$ and $\delta$. Under this condition, Eq. (7) represents that the net coupling strength can be obtained directly from the dispersion spectra. However, for the case of $\Delta_{geom} \sim \Delta_{mat}$, it leads to $\Delta' \approx 0$; thus, the frequency gap in the dispersion spectra disappears. It has been reported that this type of dispersion is referred to as crossing or weak coupling in other coupled systems [23, 32, 40]. On the other hand, for the case of $\Delta_{geom} < \Delta_{mat}$, $\Delta'$ should be imaginary, resulting in always opposite anti-crossing dispersions. Therefore, for the last two cases, the net coupling strength can be determined by the full expression of Eq. (9). Thus, both terms of $\Delta_{mat}=(\beta-\alpha)^2/2K^2$ and $\Delta_{geom}=(1+\delta\cos\phi)$ determine the net coupling strength, consequently resulting in the coupling dispersion type. In our case, $\Delta_{mat}$ is estimated to be 0.215, and thus we found the opposite anti-crossing dispersion for case-II ($\Delta_{geom} < 0.215$) and the normal dispersion for case-I ($\Delta_{geom} > 0.215$).

In earlier studies of optomechanical systems, the level attraction usually was achieved by controlling the dissipation rate (i.e. damping) of individual systems [27], i.e., using the intrinsic term $\Delta_{mat}$. However, in the present study, we explored the effect of the geometry term $\Delta_{geom}$ (i.e., both $\phi$ and $\delta$) on the dispersion type of photon-magnon coupling in the YIG/ISRR hybrid system. In this study, the good agreements between the analytical derivation/numerical calculations and experimental observations suggest that a means of controlling both $\phi$ and $\delta$ would make for great potential applications in information-processing technologies.

## IV. CONCLUSIONS

In conclusion, an experimental demonstration of the abnormal opposite anti-crossing dispersion (or level attraction) was achieved at room temperature by using a photon-magnon-coupled system that consists of a YIG film and specially designed ISRR structures in the planar geometry. The anti-crossing effects, including the dispersion type, the linewidth, and the net coupling strength of the two coupled modes, are remarkably variable and controllable by changing the relative strengths and phases of the oscillating magnetic fields generated from both the ISRR's split gap and the microstrip feeding line. The experimentally observed opposite anti-crossing and the analytically calculated abnormal anti-crossing of various dispersion types demonstrate the potential and great flexibility of photon-magnon systems for exploration of the not-yet-revealed phenomena of light-matter interaction. Although this new phenomenon requires further detailed investigation, our simple experimental design and analytical derivations could be exploited for different coupled systems.

# ACKNOWLEDGMENTS

This research was supported by the Basic Science Research Program through the National Research Foundation of Korea (NRF) funded by the Ministry of Science, ICT & Future Planning (NRF-2018R1A2A1A05078913). The Institute of Engineering Research at Seoul National University provided additional research facilities for this work.

# References

[1] H. J. Kimble, Nature, **453**, 1023 (2008).

[2] M. Wallquist, K. Hammerer, P. Rabl, M. Lukin, and P. Zoller, Phys. Scr., **T137**, 014001 (2009).

[3] Z. Xiang, S. Ashhab, J. You, and F. Nori, Rev. Mod. Phys., **85**, 623 (2013).

[4] A. Imamoglu, Phys. Rev. Lett., **102**, 083602 (2009).

[5] H. Huebl, C. W. Zollitsch, J. Lotze, F. Hocke, M. Greifenstein, A. Marx, R. Gross, and S. T. B. Goennenwein, Phys. Rev. Lett., **111**, 127003 (2013).

[6] Y. Tabuchi, S. Ishino, T. Ishikawa, R. Yamazaki, K. Usami, and Y. Nakamura, Phys. Rev. Lett., **113**, 083603 (2014).

[7] X. Zhang, C. L. Zou, L. Jiang, and H. X. Tang, Phys. Rev. Lett., **113**, 156401 (2014).

[8] M. Goryachev, W. G. Farr, D. L. Creedon, Y. Fan, M. Kostylev, and M. E. Tobar, Phys. Rev. Appl., **2,** 054002 (2014).

[9] Y. Cao, P. Yan, H. Huebl, S. T. B. Goennenwein, and G. E. W. Bauer, Phys. Rev. B, **91**, 094423 (2015).

[10] L. Bai, M. Harder, Y. P. Chen, X. Fan, J. Q. Xiao, and C.-M. Hu, Phys. Rev. Lett., **114**, 227201 (2015).

[11] X. Zhang, C. Zou, L. Jiang, and H. X. Tang. J. Appl. Phys., **119**, 023905 (2016).

[12] L. Bai, Blanchette, M. Harder, Y. P. Chen, X. Fan, J. Q. Xiao, and C.-M. Hu, IEEE Trans. Magn., **52**, 1000107(2016).

[13] B. Bhoi, T. Cliff, I. S. Maksymov, M. Kostylev, R. Aiyar, N. Venkataramani, S. Prasad, and R. L. Stamps, J. Appl. Phys., **116**, 243906 (2014).

[14] S. Kaur, B. M. Yao, J. W. Rao, Y. S. Gui, and C.-M. Hu, Appl. Phys. Lett., **109**, 032404 (2016).

[15] H. Maier-Flaig, M. Harder, R. Gross, H. Huebl, and S. T. B. Goennenwein, Phys. Rev. B, **94**, 054433 (2016).

[16] M. Harder, L. Bai, C. Match, J. Sirker, and C. M. Hu, Sci. China Phys. Mech. Astron., **59**, 117511 (2016).

[17] D. Zhang, W. Song, and G. Chai, J. Phys. D: Appl. Phys., **50**, 205003 (2017).

[18] B. Bhoi, B. Kim, J. Kim, Y-J. Cho, and S.-K. Kim, Sci. Rep., **7**, 11930 (2017).

[19] V. Castel, R. Jeunehomme, J. Ben Youssef, N. Vukadinovic, A. Manchec, F. K. Dejene, and G. E. W. Bauer, Phys. Rev. B, **96**, 064407 (2017).

[20] B. Yao, Y. S. Gui, J. W. Rao, S. Kaur, X. S. Chen, W. Lu, Y. Xiao, H. Guo, K.-P. Marzlin and C.-M. Hu, Nat. Commun., **8**, 1437 (2017).

[21] Z. J. Tay, W. T. Soh, and C. K. Ong, J. Magn. Magn. Mater., **451**, 235 (2018).

[22] N. R. Bernier, E. G. Dalla Torre, and E. Demler, Phys. Rev. Lett. 113, 065303 (2014).

[23] H. Eleuch and I. Rotter, Acta Polytech. 54, 106 (2014).

[24] A. P. Seyranian, O. N. Kirillov, and A. A. Mailybaev, J. Phys. A 38, 1723 (2005).

[25] A. Metelmann and A. A. Clerk, Phys. Rev. Lett. 112, 133904 (2014).

[26] M. Khanbekyan, H. A. M. Leymann, C. Hopfmann, A. Foerster, C. Schneider, S. Höfling, M. Kamp, J. Wiersig, and S. Reitzenstein, Phys. Rev. A 91, 043840 (2015).

[27] N. R. Bernier, L. D. Toth, A. K. Feofanov, and T. J. Kippenberg, Phys. Rev. A, **98**, 023841 (2018).

[28] In the publication process, we became aware that a similar level attraction between the magnon mode of a YIG sphere and a 3D-cavity photon mode had been experimentally demonstrated: see M. Harder, Y. Yang, B. M. Yao, C. H. Yu, J.W. Rao, Y. S. Gui, R. L. Stamps, and C.-M. Hu Phys. Rev. Lett., **121,** 137203 (2018).

[29] H. Xu, D. Mason, L. Jiang, and J. G. E. Harris, Nature (London), **537**, 80 (2016).

[30] Jan Wiersig Phys. Rev. Lett., **97**, 253901 (2006).

[31] Q. H. Song and H. Cao, Phys. Rev. Lett., **105,** 053902 (2010)

[32] S. H. G. Chang and C. Y. Sun, Opt. Express 24, 16822 (2016).

[33] K. Fang, J. Luo, A. Metelmann, M. H. Matheny, F. Marquardt, A. A. Clerk, and O. Painter, Nat. Phys., **13**, 465 (2017).

[34] V. L. Grigoryan, K. Shen, and K. Xia, Phys. Rev. B, **98**, 024406 (2018).

[35] F. Martin, Artificial Transmission lines for RF and Microwave Applications. New York, USA: John Wiley & Sons, 2015.

[36] P. Gay-Balmaz and O. J. F. Martin, J. Appl. Phys., **92**, 2929 (2002).

[37] R. Bojanic, V. Milosevic, B. Jokanovic, F. Medina-Mena, and F. Mesa, IEEE Trans. Microwave Theory Tech., **62**, 1605 (2014).

[38] J. Naqui, L. Su, J. Mata, and F. Martín, Int. J. Antennas Propag., **2015**, 792750 (2015).

[39] H. Eleuch, and I. Rotter, Phys. Rev E. **87**, 052136 (2013).

[40] M. Harder, L. Bai, P. Hyde, and C.-M. Hu, Phys. Rev. B, **95**, 214411 (2017).

**Figure Legends**

**Figure 1** Schematic drawing of experimental setup for photon-magnon coupling consisting of ISRR and YIG film in the planar geometry. The ISRR is capacitively coupled to a microstrip feeding line. In the experiment, ports 1 and 2 of the feeding line are connected to a VNA, and the static applied magnetic field $H$ is created by an electromagnet applied in the *x*-direction. Insets: dimensions of ISRRs oriented orthogonal (case-I) and parallel (case-II) to microstrip feeding line

**Figure 2** Experimentally measured $S_{21}$ power on the plane of normalized microwave angular frequency and magnetic field ($\omega/\omega_{ISRR}$ - $H/H_{cent}$ plane) of ISRR-YIG hybrid for different orientations of ISRR split-gap with respect to microstrip feeding line: (a) case-I: orthogonal; (b) case-II: parallel. The black solid lines in (a) and (b) correspond to the results of the fitting of Eq. (7) to the higher and lower branches.

**Figure 3** Calculated resonance frequencies (top) and linewidths (bottom) of photon-magnon modes for (a) $\delta = 1/2$, $\phi = 0$ (normal anti-crossing) (b) $\delta = 2$, $\phi = \pi$ (opposite anti-crossing). The dashed lines in (a) show the pure photon (green line) and magnon (orange line) modes, while the vertical dotted line represents the centre of anti-crossing. The vertical dotted line in (b) shows the coupling region in opposite anti-crossing.

**Figure 4** Analytical calculation of (a) $|S_{21}|$ profiles versus $\omega/\omega_{ISRR}$ at the center position ($H_{cent}$). The *y*-axis scale for the cases of ① - ⑥ and ⑩ - ⑪ is 10 times larger than that for ⑦ - ⑨ and ⑫ - ⑮. (b) The $|S_{21}|$ power spectra on the $\omega/\omega_{ISRR}$ - $H/H_{cent}$ plane according to both $\delta$ and $\phi$, the values being indicated by the numbers and positions (open circles) on the phase diagram shown in (c). The right column indicates the $|\Delta|$ as a function of $\phi$ for each of $\delta = 0$, 1, and 2. (c) Phase diagram of various types of anti-crossing dispersions on the $\delta$ - $\phi$ plane. The color indicates the absolute value of net coupling strength $|\Delta|$ noted by the two color bars for $\Delta_{mat} = (\beta - \alpha)^2 / 2K^2 \sim 0$. The black ($\Delta = 0$ for $\Delta_{mat} = 0$) and red ($\Delta' = 0$ for $\Delta_{mat} = 0.215$) lines correspond to the boundaries that distinguish dispersion types.

# Figures

Fig. 1

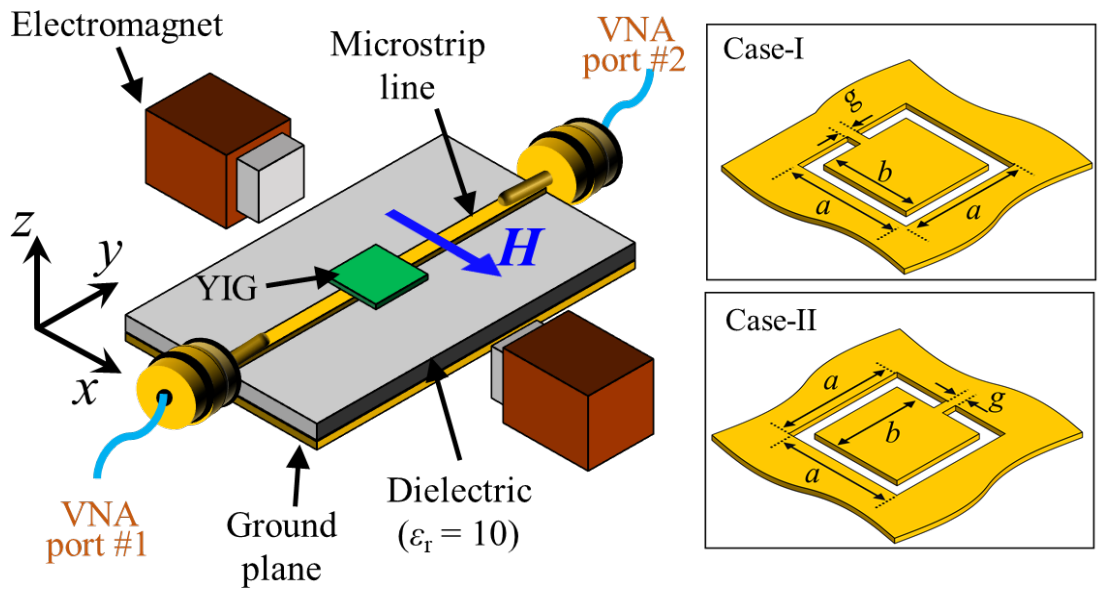

Fig. 2

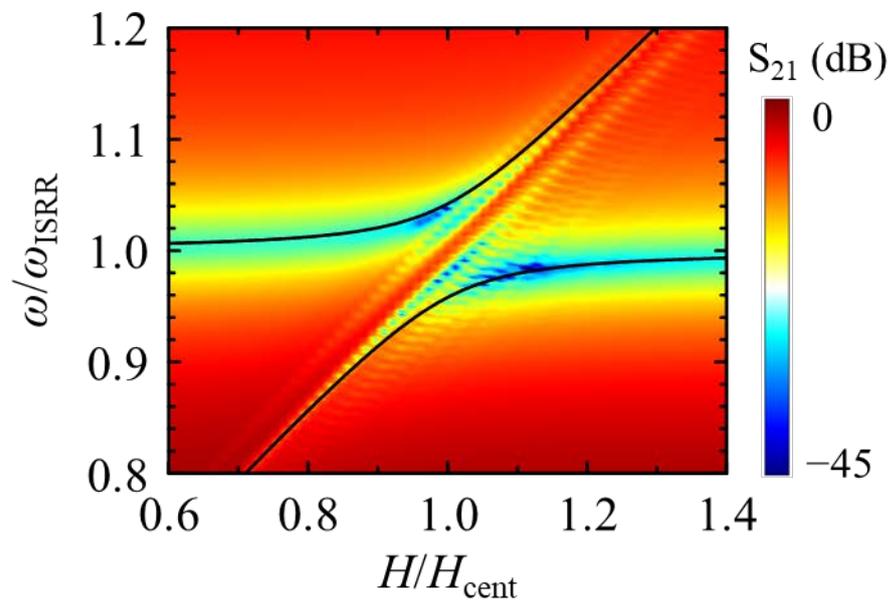

(a) Case-I

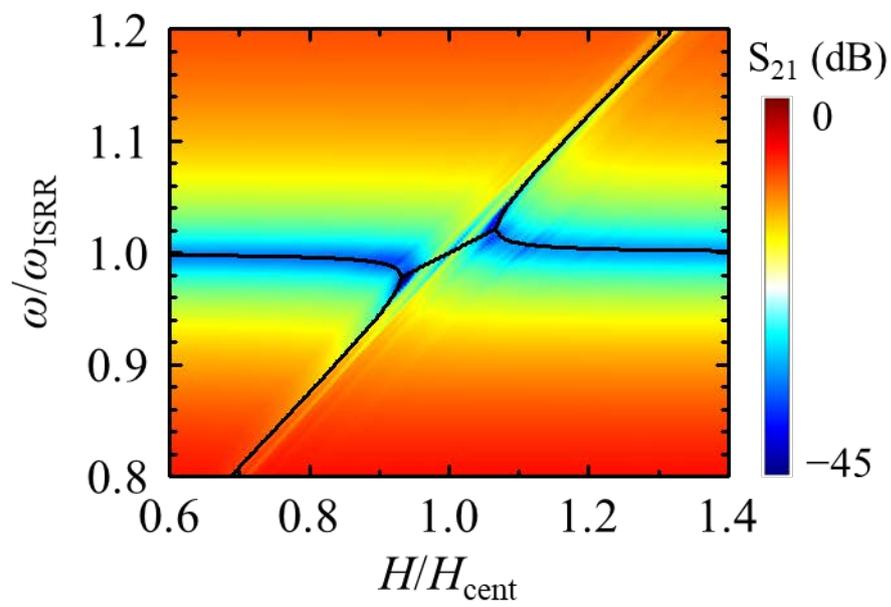

(b) Case-II

Fig. 3

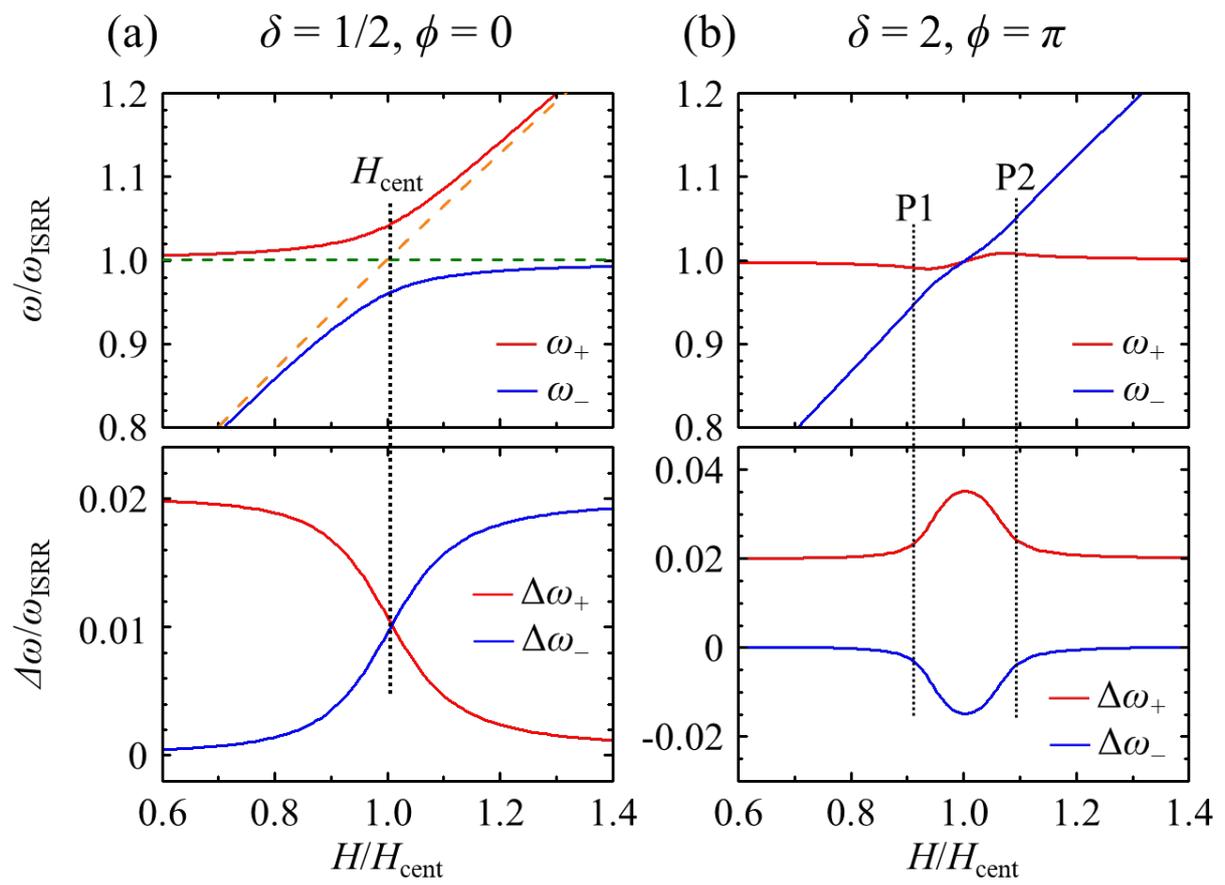

Fig. 4

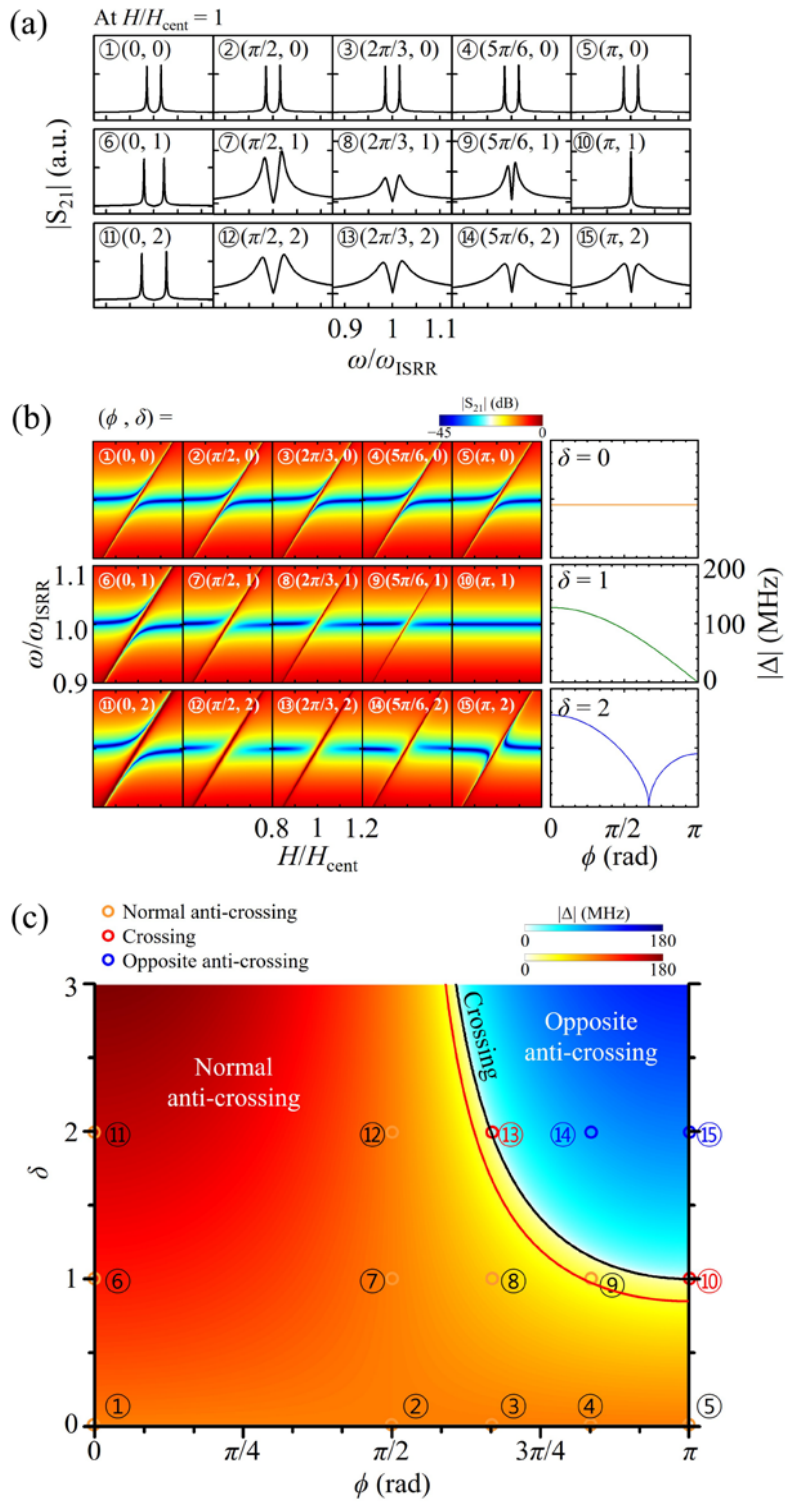

**SUPPLEMENTARY MATERIALS**

# Abnormal anti-crossing effect in photon-magnon coupling

Biswanath Bhoi, Bosung Kim, Seung-Hun Jang, Junhoe Kim, Jaehak Yang, Young-Jun Cho, and Sang-Koog Kim[a)]

*National Creative Research Initiative Center for Spin Dynamics and Spin-Wave Devices, Nanospinics Laboratory, Research Institute of Advanced Materials, Department of Materials Science and Engineering, Seoul National University, Seoul 151-744, Republic of Korea*

a) Correspondence and requests for materials should be addressed to S.-K. K

sangkoog@snu.ac.kr

## S1. Simulation results of transmission coefficient $S_{21}$ for only the ISRR structure of two different position/orientations of its split gap with respect to that of the microstrip line

Figures S1(a) and S1(b) show the amplitude and the phase of the transmission coefficient ($S_{21}$) spectra, respectively, for two different geometries of the ISRRs' split gap (case-I vs. case-II) (see Fig. 1), as obtained from electromagnetic simulations using the CST microwave studio. The resonance frequency positions for both the case-I and case-II geometries are marked by orange and green dashed lines on their amplitude and phase spectra. The case-II geometry shows a larger amplitude and phase shift in $S_{21}$ than does case-I. These contrasting results are attributable to the two cases' different orientations and positions of the ISRR's split gap with respect to the microstrip line, because both the electric and magnetic field components generated by ac currents differently affect the photon-mode excitation [SR1-SR5].

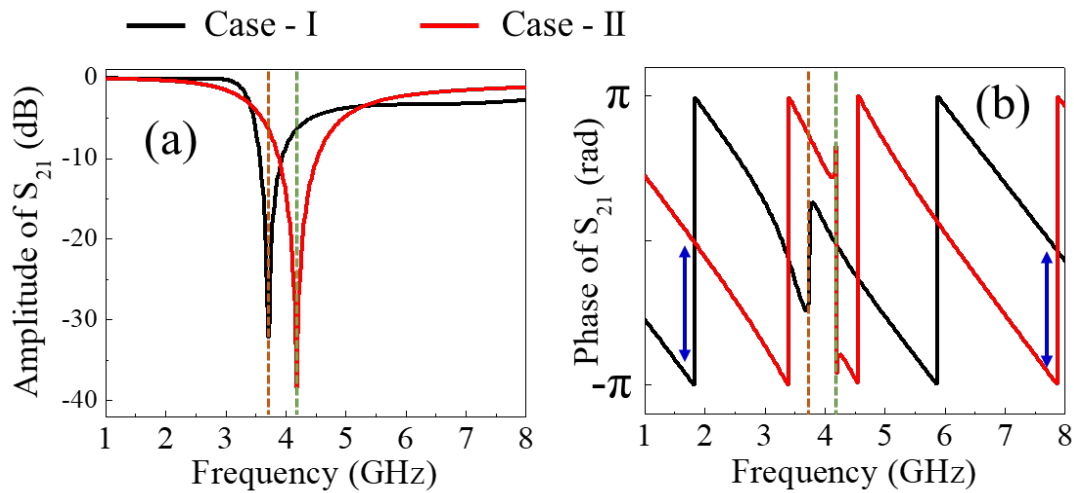

**Fig. S1.** Comparison of (a) amplitude and (b) phase of $S_{21}$ versus frequency of ac currents for ISRR structures of two different split-gap position/orientations (case-I vs. case-II) as noted in Fig. 1 (manuscript). The vertical dashed lines correspond to 3.7 and 4.1 GHz for case-I and case-II, respectively.

The numerical simulation results show that the spatial distributions of the surface current density in the ground plane (see Fig. S2) are affected by the ISRR's resonance photon modes as excited by ac currents flowing through the microstrip line.

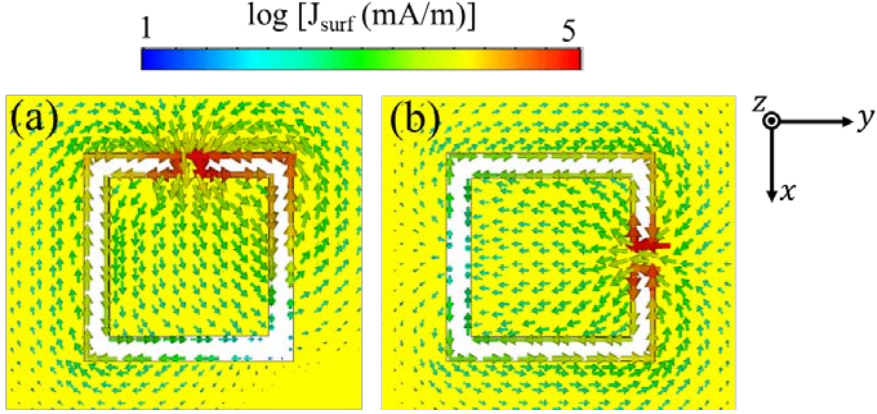

**Fig. S2**: Surface current density ($J_{surf}$) in ground plane for different orientations of split-gap with respect to microstrip feeding line: (a) case-I; (b) case-II. The arrows and colors of the arrows indicate the direction and magnitude of the surface current density, respectively. The frequency of the ac currents was 3.7 and 4.1 GHz for case-I and case-II, respectively.

**S2. Estimation of damping constant in ISRR**

The damping in the ISRR structure was obtained using $\beta = \Delta\omega_{HWHM}/\omega_{ISRR}$ [Fig. S3], where $\Delta\omega_{HWHM}$ is the half width at half maximum of the $|S_{21}|$ spectra and $\omega_{ISRR}$ is the angular resonance frequency of the corresponding ISRR structure without YIG [SR6]. The parameters $\Delta\omega_{HWHM}$ and $\omega_{ISRR}$ were estimated to 88 (80) MHz and 3.7 (4.1) GHz by fitting with a Lorentz function to the experimentally observed $|S_{21}|$ spectra for the case-I (case-II) geometry of the ISRR split-gap (see Fig. S3), consequently resulting in $\beta = 2.3(\pm 0.04)\ 10^{-2}$ and $1.9\ (\pm 0.08)\ 10^{-2}$ for case-I and case-II, respectively. Since the damping of the ISRR is relatively insensitive to the split-gap orientation, we used an average value of $\beta = 2.0 \times 10^{-2}$ over the two geometries for all of the numerical calculations of the analytical forms.

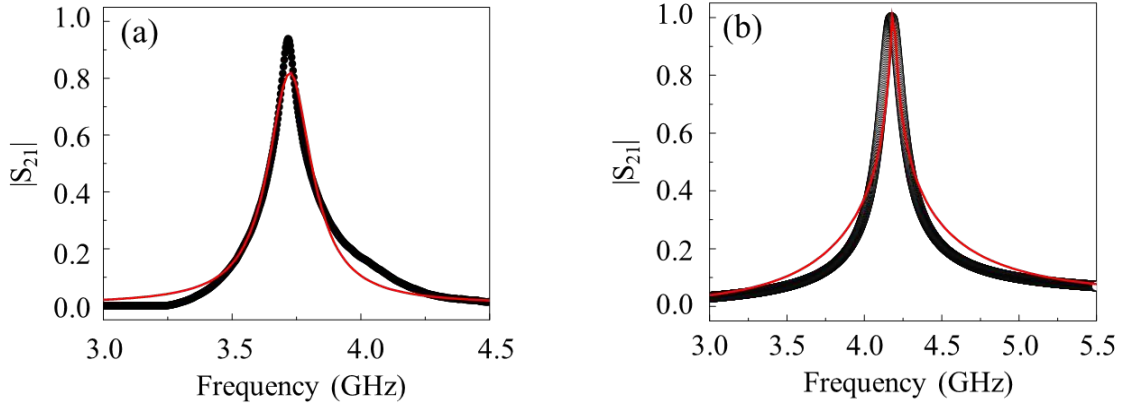

**Fig. S3.** $|S_{21}|$ spectra as function of ac current frequency for ISRRs with different split-gap orientations/positions, (a) case-I and (b) case-II. The black-dot symbols and red lines represent the experimental data and Lorentzian fitting, respectively.

## S3. Analytical forms of real and imaginary parts of $S_{21}$

The real and imaginary parts of the transmission coefficient $S_{21}$ shown by Eq. (8) in the manuscript can be rewritten as

$$\text{Re}(S_{21}) = \Gamma \frac{\omega^6(1+\alpha^2) - \omega^5\left[2\omega_r + K^2\omega_m(1+\delta\cos\phi + \alpha\delta\sin\phi)\right] + \omega^4\left[\omega_r^2 - \omega_{ISRR}^2(1+\alpha^2) + K^2\omega_r\omega_m(1+\delta\cos\phi)\right] + \omega^3(2\omega_r\omega_{ISRR}^2) - \omega^2(\omega_r^2\omega_{ISRR}^2)}{\left[-2\alpha\beta\omega^2\omega_{ISRR} + (\omega^2-\omega_{ISRR}^2)(\omega-\omega_r) - K^2\omega_m\omega^2(1+\delta\cos\phi)\right]^2 - \left[\alpha\omega(\omega^2-\omega_{ISRR}^2) + 2\beta\omega\omega_{ISRR}(\omega-\omega_r) - \delta K^2\omega_m\omega^2\sin\phi\right]^2}$$

(S1a)

$$\text{Im}(S_{21}) = \Gamma \frac{\omega^5\left[-2\beta\omega_{ISRR}(1+\alpha^2) - \alpha K^2\omega_m + \delta K^2\omega_m(\sin\phi - \alpha\cos\phi)\right] + \omega^4\omega_r(4\beta\omega_{ISRR} - \delta K^2\omega_m\sin\phi) - \omega^3(2\beta\omega_{ISRR}\omega_r^2)}{\left[-2\alpha\beta\omega^2\omega_{ISRR} + (\omega^2-\omega_{ISRR}^2)(\omega-\omega_r) - K^2\omega_m\omega^2(1+\delta\cos\phi)\right]^2 - \left[\alpha\omega(\omega^2-\omega_{ISRR}^2) + 2\beta\omega\omega_{ISRR}(\omega-\omega_r) - \delta K^2\omega_m\omega^2\sin\phi\right]^2}$$

(S1b)

## S4. Derivation of approximate solution for net coupling strength

We formulated simple analytical expressions for both types of anti-crossing dispersion in the coupling region using

$$\det \Omega = (\omega - \omega_r + i\alpha\omega_r)(\omega^2 - \omega_{ISRR}^2 + 2i\beta\omega\omega_{ISRR}) - K^2\omega_m\omega^2(1+\delta e^{i\phi}) = 0. \quad (S2)$$

Under the present experimental conditions, assuming $\alpha = 3.2\times 10^{-4}, \beta = 2.0\times 10^{-2} \ll 1$, the real part of $\Omega = 0$ becomes

$$(\omega - \omega_r)(\omega^2 - \omega_{ISRR}^2) - K^2\omega_m\omega^2(1+\delta\cos\phi) = 0. \quad (S3)$$

At $\omega \sim \omega_{ISRR}$, it turns out to be

$$\omega^2 - (\omega_r + \omega_{ISRR})\omega + \left(\omega_r\omega_{ISRR} - \frac{1}{2}K^2\omega_m\omega_{ISRR}(1+\delta\cos\phi)\right) = 0. \quad (S4)$$

Finally, we have higher ($\omega_+$) and lower ($\omega_-$) frequencies of the two coupled modes, as given by

$$\omega_{\pm} = \frac{1}{2}\left[(\omega_r + \omega_{ISRR}) \pm \sqrt{(\omega_r - \omega_{ISRR})^2 + 2K^2\omega_m\omega_{ISRR}(1+\delta\cos\phi)}\right]. \quad (S5)$$

Next, we define the net coupling strength $\Delta \equiv \frac{1}{2}(\omega_+ - \omega_-)/2\pi$ (in Hz units) as half of the gap in frequency between the upper and lower branches at $H = H_{\text{cent}}$ (where $\omega_r = \omega_{ISRR}$). Then, the net coupling strength turns out to be

$$\Delta = \frac{1}{4\pi}\sqrt{2K^2\omega_m\omega_{ISRR}(1+\delta\cos\phi)}. \quad (S6)$$

Finally, the resonance angular frequency of the two coupled modes are simplified in terms of $\Delta$ as

$$\omega_{\pm} = \frac{1}{2}\left[(\omega_r + \omega_{ISRR}) \pm \sqrt{(\omega_r - \omega_{ISRR})^2 + (4\pi\Delta)^2}\right]. \quad (S7)$$

## S5. Estimation of coupling constant *K*

From the present experimental data, we obtained the net coupling strength $\Delta = 90$ MHz for the case-I geometry, which value is exactly equal to a coupling strength ($g_{eff}/2\pi = \sqrt{2K^2\omega_m\omega_{ISRR}}/4\pi$) previously observed (see SR7). From the relation of $\sqrt{2K^2\omega_m\omega_{ISRR}}/4\pi = 90$ along with the experimentally observed values of $\omega_m = \gamma\mu_0 M_s = 4.8$ GHz and $\omega_{ISRR}/2\pi = 3.7$ GHz (case-I) and $\omega_{ISRR}/2\pi = 4.1$ GHz (case-II), we obtained $K = 0.030$ for case-I and $K = 0.029$ for case-II. Since $K$ is not very sensitive to the split-gap orientation/position, we used the value of $K = 0.03$ for all of the numerical calculations.

## S6. Numerical calculation of net coupling strength $|\Delta|$ as function of $\phi$ for different values of $\delta = 0, 1,$ and 2

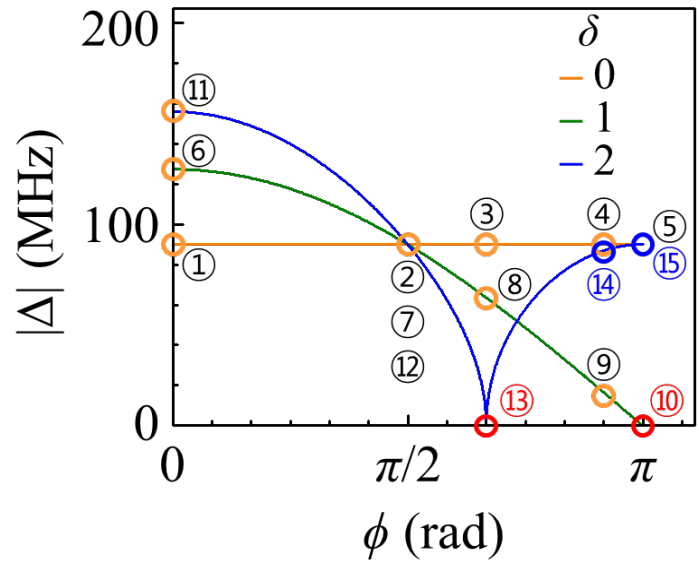

**Fig. S4.** Absolute value of net coupling strength $|\Delta|$ as function of $\phi$ for different values of $\delta = 0, 1,$ and 2. The numbers with open circles indicate the positions of $\phi$ and $\delta$ on the phase diagram shown in Fig. 4(c), which corresponds to the $|S_{21}|$ power on the $\omega/\omega_{ISRR}$ - $H/H_{cent}$ plane plotted in Fig. 4(b).

## S7. Relation among dispersion gap, coupling strength and damping

In order to understand the relation among the frequency gap between the coupled modes at the anti-crossing center, the net coupling strength and the material's damping constant, the determinant of $\Omega$ was solved to obtain its complex eigenvalues $E_\pm = \omega_\pm - i\Delta\omega_\pm$. The determinant of $\Omega$ was simplified as

$$(\omega-\omega_r+i\alpha\omega_r)\left[(\omega-\omega_{ISRR})(\omega+\omega_{ISRR})+2i\beta\omega\omega_{ISRR}\right]-K^2\omega_m\omega^2\left[1+\delta e^{i\phi}\right]=0. \quad (S8)$$

At $\omega \sim \omega_{ISRR}$, Eq. (S8) is rewritten as

$$(\omega-\tilde{\omega}_r)(\omega-\tilde{\omega}_{ISRR})-\frac{1}{2}K^2\omega_m\omega_{ISRR}\left[1+\delta e^{i\phi}\right]=0, \quad (S9)$$

where $\tilde{\omega}_r = \omega_r - i\alpha\omega_r$ and $\tilde{\omega}_{ISRR} = \omega_{ISRR} - i\beta\omega_{ISRR}$. Then, the solution of Eq. (S9) becomes

$$E_\pm = \frac{1}{2}\left[(\tilde{\omega}_r+\tilde{\omega}_{ISRR})\pm\sqrt{(\tilde{\omega}_r-\tilde{\omega}_{ISRR})^2+2K^2\omega_m\omega_{ISRR}(1+\delta e^{i\phi})}\right]. \quad (S10)$$

The energy difference between the two eigenmodes is

$$(E_+ - E_-)^2 = (\tilde{\omega}_r-\tilde{\omega}_{ISRR})^2 + 2K^2\omega_m\omega_{ISRR}(1+\delta e^{i\phi}). \quad (S11)$$

Inserting $\tilde{\omega}_r = \omega_r - i\alpha\omega_r$ and $\tilde{\omega}_{ISRR} = \omega_{ISRR} - i\beta\omega_{ISRR}$ into Eq. (S11) at $\omega_r = \omega_{ISRR}$, the real part becomes

$$\left(\frac{\omega_+ - \omega_-}{\omega_{ISRR}}\right)^2 = \left(\frac{\Delta\omega_+ - \Delta\omega_-}{\omega_{ISRR}}\right)^2 - (\beta-\alpha)^2 + 2K^2\frac{\omega_m}{\omega_{ISRR}}(1+\delta\cos\phi). \quad (S12)$$

Thus, the frequency gap $(\omega_+ - \omega_-)/2\pi$ at the anti-crossing center can be determined from Eq. (S12). The net coupling strength $\Delta \equiv \frac{1}{2}(\omega_+ - \omega_-)/2\pi$ (in Hz units) for non-damping cases was defined as half of the gap in frequency between the upper and lower branches at $H = H_{\text{cent}}$. However, the modified net coupling strength for damping cases, under the condition of $(\Delta\omega_+ - \Delta\omega_-)/\omega_{ISRR} \sim 0$, is now expressed as

$$\Delta' = \frac{1}{2}\left(\frac{\omega_+ - \omega_-}{2\pi}\right) = \frac{\sqrt{2}K}{4\pi}\sqrt{\omega_m\omega_{ISRR}(1+\delta\cos\phi)-\omega_{ISRR}^2(\beta-\alpha)^2/2K^2}.$$

(S13)

## S8. Phase diagrams of coupling dispersion types on ($\delta$ - $\phi$) plane for indicated different values of $\Delta_{mat} = (\beta - \alpha)^2 / 2K^2$

In order to examine the effect of $\Delta_{mat} = (\beta - \alpha)^2 / 2K^2$ on the anti-crossing phase diagram, we contour-plotted $\Delta'$ for different values of $\Delta_{mat} = 0.1, 0.3, 0.5,$ and $0.9$, as shown in Fig. S5. Here the line of $\Delta' = 0$ represents the boundary that distinguishes the opposite anti-crossing dispersion from the other types. By increasing the values of $\Delta_{mat}$, the opposite anti-crossing region (blue colors) expands towards lower values of $\phi$ and $\delta$.

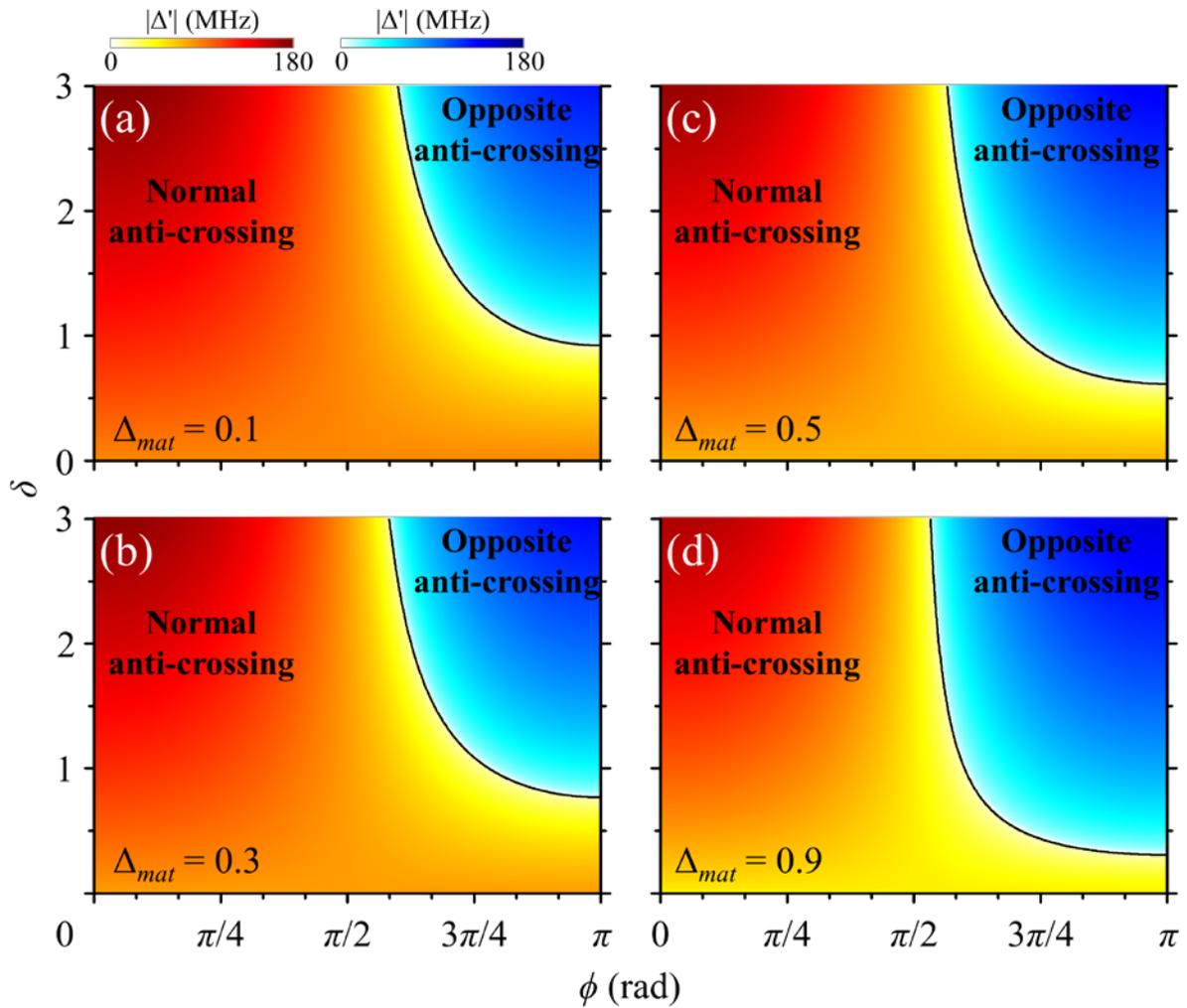

**Fig. S5.** Numerical calculation of phase diagram of anti-crossing dispersion on $\delta$ - $\phi$ plane. The color indicates the absolute value of net coupling strength $\Delta'$ as noted by each of the two color bars. The black lines correspond to the boundaries ($\Delta' = 0$) between the normal and opposite anti-crossing types for (a) $\Delta_{mat} = 0.1$, (b) 0.3, (c) 0.5, and (d) 0.9.

# Supplementary References

[SR1] P. Gay-Balmaz and O. J. F. Martin, J. Appl. Phys. 92, 2929 (2002).

[SR2] I. O. Mirza, S. Shi, and D. W. Prather, Optics Express. 17, 5089 (2009).

[SR3] N. Katsarakis *et al*., Appl. Phys. Lett., 84, 2943 (2004).

[SR4] R. Bojanic, *et al*., IEEE Trans. Microw. Theory Tech. 62, 1605 (2014).

[SR5] F. Martin, Artificial Transmission lines for RF and Microwave Applications. New York, USA: John Wiley & Sons, 2015.

[SR6]. V. Castel, R. Jeunehomme, J. B. Youssef, N. Vukadinovic, A. Manchec, F. Dejene, and G. E. W. Bauer, Phys. Rev. B 96, 064407 (2017).

[SR7] B. Bhoi, B. Kim, J. Kim, Y.-J. Cho, and S.-K. Kim, Sci. Rep. 7, 11930 (2017).